# Breakdown of 1 D water wires inside Charged Carbon Nanotubes

## Shashank Pant*

Molecules Biophysics Unit, Indian Institute of Sciences, Bangalore 560012

**ABSTRACT**: Using Molecular Dynamics approach we investigated the structure, dynamics of water confined inside pristine and charged 6,6 carbon nanotubes (CNTs). This study reports the breakdown of 1D water wires and the emergence of triangular faced water on incorporating charges in 6,6 CNTs. Incorporation of charges results in high potential barriers to flipping of water molecules due to the formation of large number of hydrogen bonds. The PMF analyses show the presence of ~2 kcal/mol barrier for the movement of water inside pristine CNT and almost negligible barrier in charged CNTs.

**Introduction**

Nanoscopic pores of CNTs (Carbon Nanotube) have gained much attention owing to their potential applications[1-3] in the area of water-desalination membranes,[4] lab-on-chip devices[5] and integrated circuits.[6] At a fundamental level the structure of water confined inside carbon nanotubes is of great importance because of the presence of ice like water inside CNTs at room temperature.[7, 8] Another important feature is the formation of 1D- water wires inside (6, 6) pristine CNTs which is of great significance in biology as well as desalination membranes.[9, 10] Moreover, CNTs can be used as a prototype to study molecular mechanism of water transport through complex biological systems such as aquaporin.[11] Earlier Hummer *et al*[12] studied the movement of water through (6, 6) CNTs using molecular dynamics simulations and observed the presence of water for 66ns long simulations. On lowering the van der Waals (vdW) interaction parameters between the nanotube and water the authors observed a wet to dry transition, which show the movement of water through CNTs strongly depends on the vdW interactions between the nanotubes and water molecules.

There are several theoretical models[12, 13] which describe the permeation of water through nanotube. Simulations have provided invaluable insights about the structure, dynamics and thermodynamics[14, 15] of water confined inside pristine nanotubes. A lot of work has been done [12, 16-18] to study the behavior of water inside nanotubes using various water models such as SPC, SPC/E, TIP series, coarse grained models and they are in good agreement with each other. However, experimentally it is difficult to work with pristine CNTs since the separation and manipulation of CNTs is impeded due to its insoluble nature in various solvents [19]. Also, studying functionalized CNT is more desirable as they can be treated as a prototype for several complex systems with a range of heterogeneity in charge distribution and chemical character.

Thus owning to the improvement in its processability [20, 21] and solubility the functionalization of CNT is desirable. Successful functionalization of CNTs has been achieved through chemically [22] modifying the walls but this might lead to the changes in electronic and structural properties. Thus before conducting these experiments this system deserve theoretical investigation. Molecular Dynamics (MD) studies have been conducted to study the properties of water in CNTs functionalized at the terminal end [23], which showed selectivity towards cation and anions. Zheng and co-workers [24] incorporated –COOH groups inside the CNTs and studied the dynamics of water-methanol system. They showed the diffusion of water-methanol mixture is slower inside the hydrophilic tubes compared to the tube with hydrophobic character. There are a few studies where functionalization was limited to the end groups of nanotubes and salination-desalination was studied [25, 26].

However, despite of years of research work in CNT water permeation, we still lack the complete knowledge of the effect of functionalization on the confined water. For example, a) what is the effect on 1D water wires seen inside 6,6 pristine CNTs on introducing hydrophilic (charged) groups? b) Are 1D water wires seen only due to hydrophobic nature CNTs? How patterning of functionalized / charged groups affects confined water? To shed lights on these questions we performed molecular dynamics simulations of functionalized carbon nanotubes in water bath. The functionalization is carried out by incorporating charges on the wall of CNT in different configurations without disturbing the size of carbon atoms.

**Methods**

All MD simulations were carried out using NAMD2 [27], in NVT ensemble followed by NPT ensemble with 2fs time step. In pre-equilibrated water box (TIP3P) 6,6 CNT of diameter 8.1 Å, length 13.1Å was inserted, and all the overlapping molecules were removed. The box was 33 X 32 X 33 Å$^3$ in volume. Periodic boundary conditions were applied in all the directions and electrostatic forces were calculated using Particle Mesh Ewald Method [28]. The vdW interactions were calculated with a cutoff of 8-10 Å. We employed langevin dynamics to control the temperature of the system at 300 K and langevin piston method was used to control the pressure at 1 atm. The carbon atoms of CNT were fixed during the entire simulation period. To construct different functionalized CNTs we assigned charges to the carbon atoms as mentioned in Table I. Total simulation time was upto 24 ns for each CNT. In order to neutralize the system counter ions were added to the simulation box. The net vdW potential between CNT and water was modeled by:

$$U_{CW} = \epsilon_{CW} \sum_{i=1}^{N_W} \sum_{j=1}^{N_C} \left[ \left( \frac{\sigma_{CW}}{r_{ij}} \right)^{12} - 2 \left( \frac{\sigma_{CW}}{r_{ij}} \right)^6 \right]$$

Where $N_c$ and $N_w$ are the number of carbon atoms in CNT and water respectively, $r_{ij}$ is the center-to-center distance between water and carbon atoms of CNT. Cross terms of $\sigma$ and $\epsilon$ were calculated using Lorentz-Berthelot rules. Table 1 summarizes the vdW and charged parameters used in the force field:

Table 1.

| Name | | (R/2) (Å) | $\varepsilon \left( \frac{kcal}{mol} \right)$ | q (e) |
|---|---|---|---|---|
| Type1 | C | 1.9924 | -0.07 | 0 |
| | H | 0.2245 | -0.046 | |

| Type | Atom | | | |
|------|------|--------|---------|----------------|
| | O | 1.7682 | -0.1521 | |
| | | | | |
| Type2 | C | 1.9924 | -0.07 | +0.550 / -0.500 |
| | H | 0.2245 | -0.046 | |
| | O | 1.7682 | -0.1521 | |
| | | | | |
| Type3 | C | 1.6418 | -0.0277 | 0 |
| | H | 0.2245 | -0.046 | |
| | O | 1.7682 | -0.1521 | |

charges are assigned according to C, =O of –COOH group [29]; R = $\sigma_{CW}$

We also ran simulations by distributing charges on carbon atoms of CNT in the form of stripes. The net charges used in this study are as follow: -6, -12, -18, -24 esu respectively as shown in Figure 1.

**Results and Discussion**

MD simulation was carried out for both pristine (hydrophobic) and functionalized / charged (hydrophilic) CNTs.

1. **Variation in confined water molecules**

Although, CNT is hydrophobic in nature its channel gets rapidly filled up with water and remains occupied with around 4-5 water molecules. This is consistent with previously reported computational, theoretical [12, 30] and experimental studies [31]. Figure 2(a) show the incorporation of charges on CNT increases the number of confined water molecules without showing any wetting/dewetting transitions. We studied the effect of charge on the average number of confined water by calculating the change in the average number of confined water molecules with charges. The average number of confined water molecules increase exponentially with the charge percentage as seen in figure 2(b). This increase in average number of confined water molecules is in correspondence with earlier computational study [32] and suggests that the electrostatic interactions between the CNT and water play a dominant role compared to the vDW interactions [13] during the filling process.

In-order to see wetting/dewetting transition and to quantify hydrophobicity, [12] we decreased the vdW attractive interaction between water and CNTs and observed number of wetting / dewetting transitions as shown in figure 3(a). Also, simulations were performed at different temperatures (280, 290, 300, 320 K) for pristine CNT to understand the effect of temperature on the number of confined water molecules. Figure 3(b) shows there is hardly any change in the average number of confined water molecules within the studied range of temperature, which is in good approximation with Chopra et al [16].

## 2. Structure of water inside & around CNTs

To estimate the structural ordering of water molecules inside CNT, we calculated pair correlation function of the water molecules along the axis of CNT (Z axis).

$$g(z) = \frac{1}{N} \sum_{i=1}^{N} \sum_{j=1, j \neq i}^{N} < \delta \left( z - z_{ij} \right) >$$

Where N is the number of confined water molecules, $z_{ij}$ is the axial separation between $i^{th}$ and $j^{th}$ particle and, angular brackets indicate time average. Figure 4(a), show the presence of well separated peaks in the pair correlation function of confined water indicating solid like ordering of water inside the CNT even at 300 K, is in good correspondence with [16, 33]. The nearest neighbor distance between the water molecules was estimated to be around 2.6 Å, as reported in previous study [12]. A closer look reveals decrease in ordering of confined water molecules with the incorporation of charges on CNT, which also supports the presence of ordered water around the hydrophobic solutes [34-38]. We further observe the effect of temperature on the structure of water along the axis of CNT as shown in figure 4(b). Even at 320 K, there is hardly any change in the structure of confined water and it exists in solid like form with peaks present at almost regular intervals. Further we calculated the radial density profile of water in and around pristine and charged CNTs as shown in figure 5. The radial density profile of pristine show a broad / flat region of water density inside CNT which bifurcates into two separate peaks in charged CNTs. To find the reason behind this bifurcation we observe the last snapshot from the MD simulation Figure 6(a), which show the presence of triangular faced confined water at the charged end of (6,6) CNT that tapers to the other end of tube. To best our knowledge this is the first study which shows the presence of triangular faced water inside charged 6,6 armchair CNT. To understand the reason behind the tapering of water we distributed these charged atoms at the center of CNT (keeping the number of charge atom fixed) and observed triangular network of water throughout the CNT channel figure 6(b) which suggest that the electrostatic interactions can be used to tune the properties and structure of confined water. The radial density profile of water outside CNT show the movement of water density away from CNT surface in the case of pristine CNT, while water density moves significantly closer to the surface in charged CNTs. The distribution of charges in the form of stripes keeping the total charge fixed to -18.0 esu also affects the water structure as shown on Figure 6(c). Table 2 and figure 1 shows the distribution of charges on CNT:

TABLE 2

| Name | Pattern |
|------|---------|
| Stripe 1 | Charge on 36 C atoms at one end. |
| Stripe 2 | Charge on 18 C atoms at both the ends. |
| Stripe 3 | Charge on 12 C atoms at 3 locations (alternate). |
| Stripe 4 | Charge on 9 C atoms at 4 locations (alternate). |

Close analysis of Figure 6(c) show the presence of water density outside on the walls of CNT that increases as the number of stripes were increased to 4. This increase in peak shows the wetting of CNT i.e. charged CNTs can be solubilized as the numbers of stripes are increased. There is also significant change in the structure of confined water, which show the presence of bifurcated peaks for stripes 1, 2 and 3. In the case of stripe 4 we observe all the water molecules are along one side of CNT and close analysis of MD snapshot (not shown) shows the presence of counter ions on the other side of CNT. Such type of systems can be effectively used to study the behavior of water movement through channel proteins like aquaporin and the presence of ions can work as the charged constriction regions seen in aquaporin [39].

### 3. Permeation of water through CNT channel

A permeation event is considered when water molecule crosses from one end to the other end of CNT. We analyzed the permeation of water molecules through pristine and charged CNT. Figure 7(a) shows decrease in the number of permeation events with the increase in charge percentage. In pristine CNT, we observe 4.5 water permeation events per ns, which is in good correspondence with Zhu et al [11] but on introduction of charges, number of permeation events decrease significantly to around 3 per ns which is twice as observed in aquaporin [39]. This shows the introduction of charge blocks the CNT channel and further hinders the movement of particles. We can use these ideas to design functionalized CNTs and use them as a model system for studying biological water channels and on-off gating mechanisms [39]. The permeation of water molecules varies almost linearly with the temperature as shown in Figure 7(b), which can be used as a handle to tune the number of permeation events. The distribution of charges in the form of stripes hardly affects the water flow as shown in figure 6(c), which suggests the flow of water is affected by the net charge not by the distribution of charge on the CNT.

### 4. Density fluctuations in confined water

In-order to gain the macroscopic understanding of confined space we calculated the local density fluctuations in the confined water. The presence of low-density vapor like region in the confined space increases with the increase in the hydrophobicity of the CNT. To quantify the fluctuations in the confined region we determined the fluctuations in the water occupancy numbers around the probe particles of 3 Å diameters, which were placed inside the CNT. Figure 8 shows the variance in water density fluctuations normalized over the number of water molecules, as a function of charged and pristine CNT. In the region of macroscopic limit, the ratio of variance and average approaches $\rho k_B T \chi_T$, where $\chi_T$ is the isothermal compressibility. Figure 8 show 2 horizontal lines as the limiting cases of fluctuations in bulk water and ideal gas. As the hydrophobicity of CNT is increased (pristine CNT with reduced VdW interactions) $\kappa$ increase to almost 5 times the value of bulk water. This large variance in the fluctuation suggests that the water near hydrophobic CNT is more compressible owning to the wet-dry transitions. Large compressibility was earlier reported for water around large solutes [40] and around hydrophobic plates [41].

$$\kappa = \frac{(<N^2> - <N>)}{<N>}$$

### 5. Potential of Mean Force

To understand the water transport inside CNT, we analyzed PMF for water permeation in pristine and charged CNT using the following relation.

$$PMF(z) = -k_B T ln(\frac{\rho(z)}{\rho_o})$$

where is the density of water along the axis of CNT (Z axis in this study) and is the bulk water density. In figure 9, we observe PMF profile of water permeating through the two different CNTs (black curve for pristine CNT and red for charged CNT). The free energy profile shows the presence of wave like water distribution inside the CNT and valleys at the end, which signifies distinct water density at the end. The wavy nature of PMF profile indicates ordering of water molecules due to tight hydrogen bonding network, [42] due to solid like nature of confined water. We observe the free energy barrier for water moving inside pristine CNT was ~2kcal/mol while no such barrier was observed for charge CNT. Also there was not well distinct wave like water density inside charged CNTs which supports our point of the presence of ordered water inside pristine CNT is due to its hydrophobic nature.

### 6. Dipole Moment

To understand the differences in water-water interactions for different CNT systems, we examined the orientation of average dipole moment of water with the axis of CNT shown in figure 10(a). The dipole interaction between water molecules plays an important role in the behavior of single-file water chains [43]. For pristine CNT we observe $\theta$ falls in two ranges, $15^{\circ} < \theta < 50^{\circ}$ and $120^{\circ} < \theta < 150^{\circ}$ which clearly shows a two state phenomenon of orientation distribution and is in good correspondence with the earlier works [43, 44]. A flip is defined as $\theta$ passing through $90^{\circ}$ and with the incorporation of charges on CNT we observe a drastic change in the population of the two states, for CNT with 12 charged atoms i.e. -6.0 esu at the terminal ends we observe increase in the population of $15^{\circ} < \theta < 50^{\circ}$ and corresponding decrease in the other. Further on increasing the number of charges we hardly observe any flipping events during the entire production runs and the dipole moment shift towards $90^{\circ}$ i.e. perpendicular to the axis of CNT. This is because the presence of electrostatic interactions between the charged CNT and the hydrogen atoms of water shown in figure 11 changes the direction of dipole moment. Similar decrease in number of flipping events was observed for CNTs under electric field [32]. A close analysis of peaks reveals their symmetric nature with respect to $90^{\circ}$ for pristine CNT and the incorporation of charges makes it asymmetric (for charge =12), which is again in good correspondence with Vaitheeswaran's theoretical prediction [45]. Generally flip is governed by the potential energy barrier against flipping due to the confinement of water along with the hydrogen bond energy. In order to explain, the no flipping events after incorporating charges on CNT, we calculated the average number of hydrogen bonds formed as shown in Figure 10© which suggests linear increase in the average number of hydrogen bonds with the increase in the charge %, thereby increasing the potential barrier for flipping of water molecules in the channel. Further we analyzed the effect of distributing charges in the form of stripes on the orientation of dipole moment with respect to the axis of CNT as shown in figure 10(c). The result shows no flipping events but increasing the number of stripes shift $\theta$ towards 90º.

**Conclusion**

In the present study we have used molecular dynamics approach to investigate the properties of confined TIP3P water inside pristine and functionalized/charged carbon nanotubes. We calculated the structural, dynamic properties of confined water by changing the charge percentage, distribution pattern and effect of temperature. With the increase in the charge % we observe for the first time the presence of triangular faced confined water inside charged 6,6 CNTs. This observation suggests that the presence of 1D water wires is due to the hydrophobic nature of CNT and not due to the confined space. We observe a drastic increase in the average number of confined water molecules with the increase in the charge %, also the average number of water molecules fluctuates around 3±1 water molecules on changing the distribution of charged atoms. On decreasing the VdW interactions between CNT and water we observe the presence of wetting/dewetting transitions. Structure of confined water along the axis of CNT show the presence of solid like ordering of water inside pristine CNT, which decreases in the charged CNTs. The density profile of water in and around CNT show the movement of water away from the pristine CNT surface which suggest hydrophobic nature of CNT and also the presence of charge affects the density of confined water. Permeation analysis of water molecules show a drastic decrease in number of permeation events with the incorporation of charges, which show the partial blocking of CNT channel with the incorporation of charges. Further we observe the patterning of charges in the form of stripes hardly change the number of permeation events. The hydrophobic character of CNTs was gauged by calculating compressibility of water inside CNT channel, which decreases to around (~5 times) in charged CNTs showing bulk water like behavior. PMF calculation shows the presence of ~2-kcal/mol barrier in pristine CNT for the movement of water inside the channel while no such barrier was observed for charged CNTs. The dipole moment orientation analysis shows the flipping of water dipole with respect to CNT axis in pristine CNT and such flipping events were clearly absent in charged CNTs. This flipping of water is attributed to the presence of lesser number of hydrogen bonds in pristine CNTs compared to charge ones. This study suggests that the patterning of charged atoms on the surface of nanotubes in the form of stripes hardly affect the dynamical properties of confined water but it does affect the structure of water.

AUTHOR INFORMATION

**Corresponding Author**

shashankpant.lko@gmail.com

**ACKNOWLEDGMENT**

We would like to thank Dr. Anand Srivastava for his comments and critically reading the manuscript. Thanks are due to Arjun Cluster facility at Molecule biophysics Unit, Indian Institute of Sciences Bangalore.

ABBREVIATIONS

CNT, Carbon Nanotubes

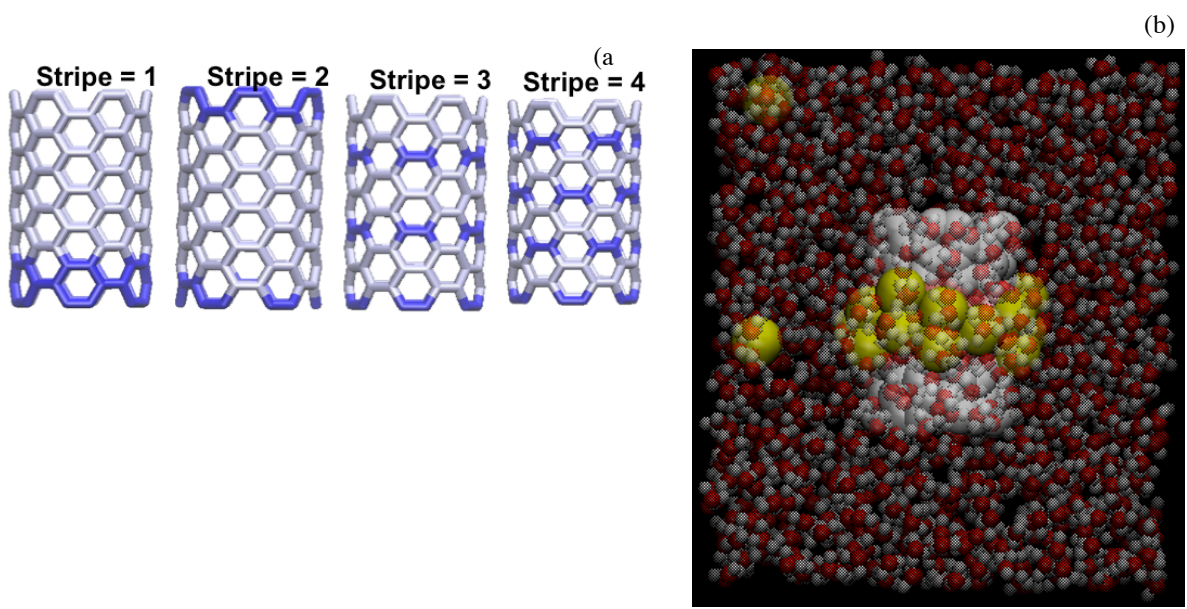

Fig. 1 (a) Show the distribution of charges on CNT in the form of stripes, (b) Snapshot of simulation box (33X32X33) Å$^3$. Yellow colored for counter ions, red for oxygen, white for hydrogen and CNT at the center.

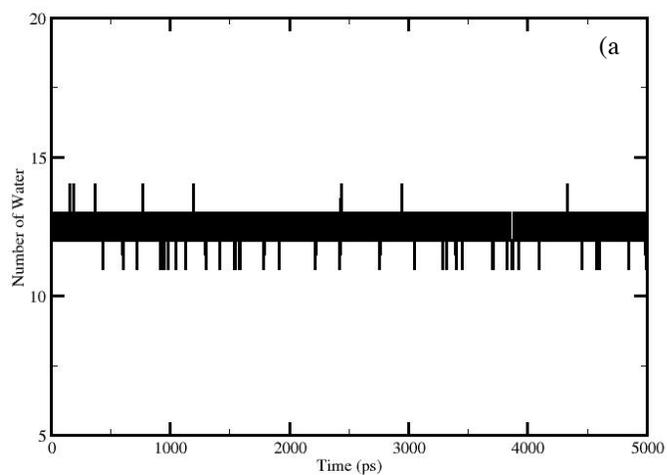

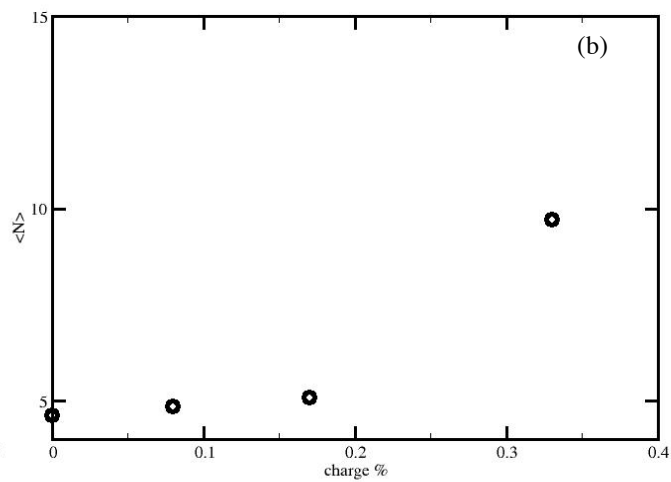

Fig. 2 (a) Show the fluctuation in the number of confined water molecules inside charged CNT (maximum charge density) (b) Variation in the average number of confined water molecules with the charge. Charge % 0 corresponds to pristine CNT.

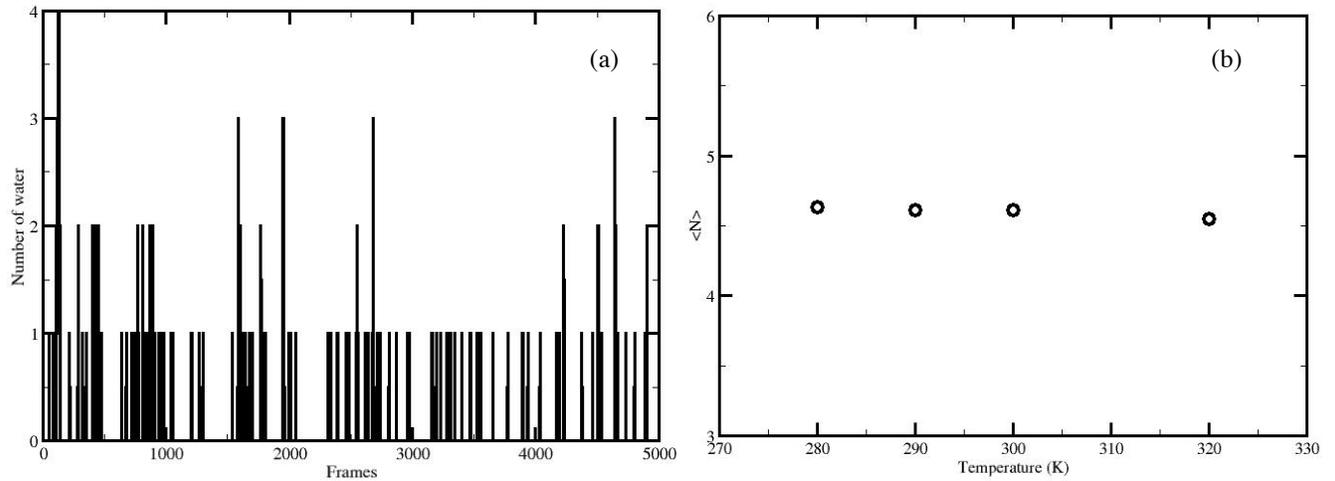

Fig. 3 (a) Show the wetting/dewetting transitions on decreasing the vdW interaction between CNT and water molecules, (b) Show the variation in the number of confined water molecules with the change in the temperature.

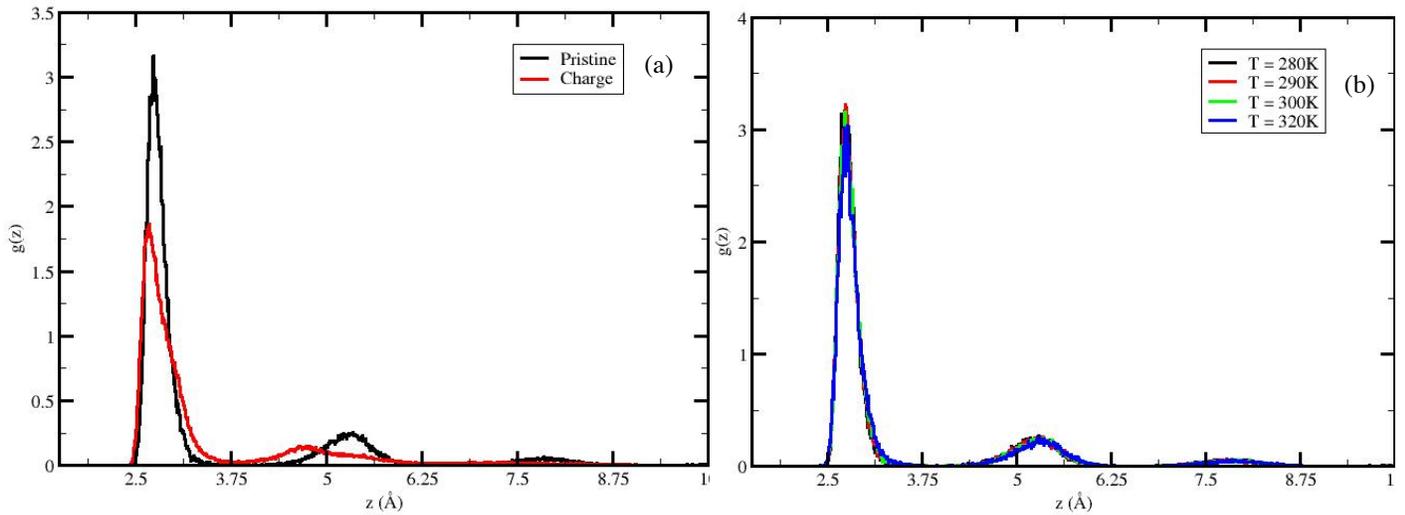

Fig. 4 (a) Radial distribution curves of the confined water along the axis of pristine and charged CNT, (b) Variation of radial distribution curves with the temperature.

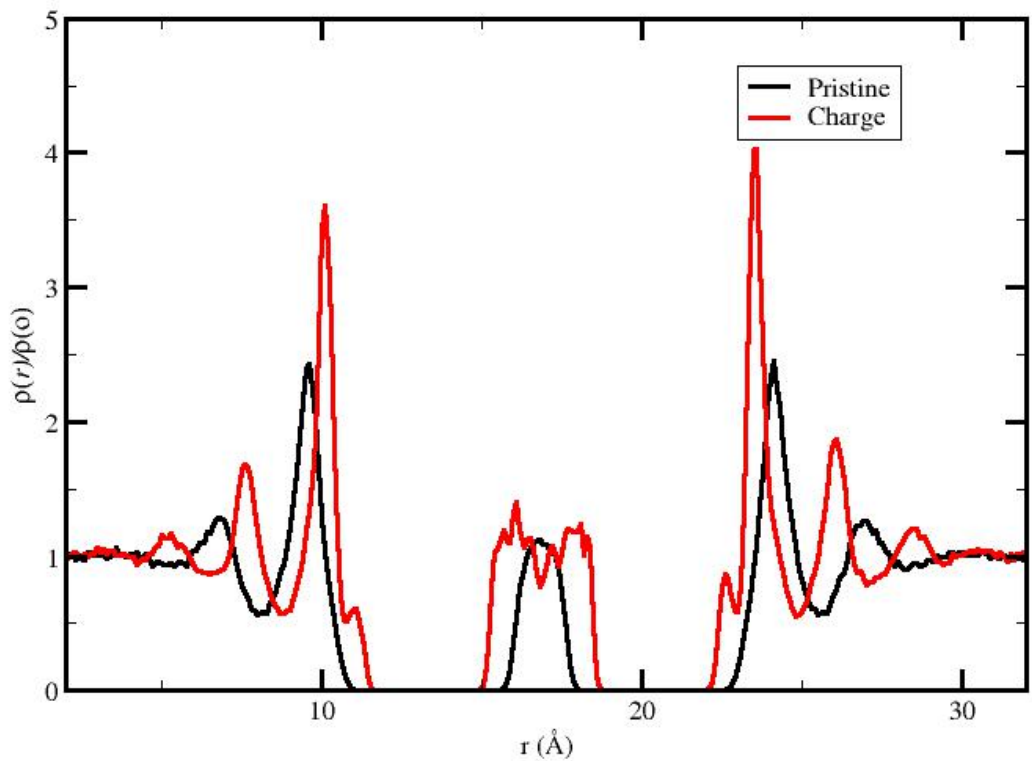

Fig. 5 Density profile of water inside and outside the pristine and charged CNTs.

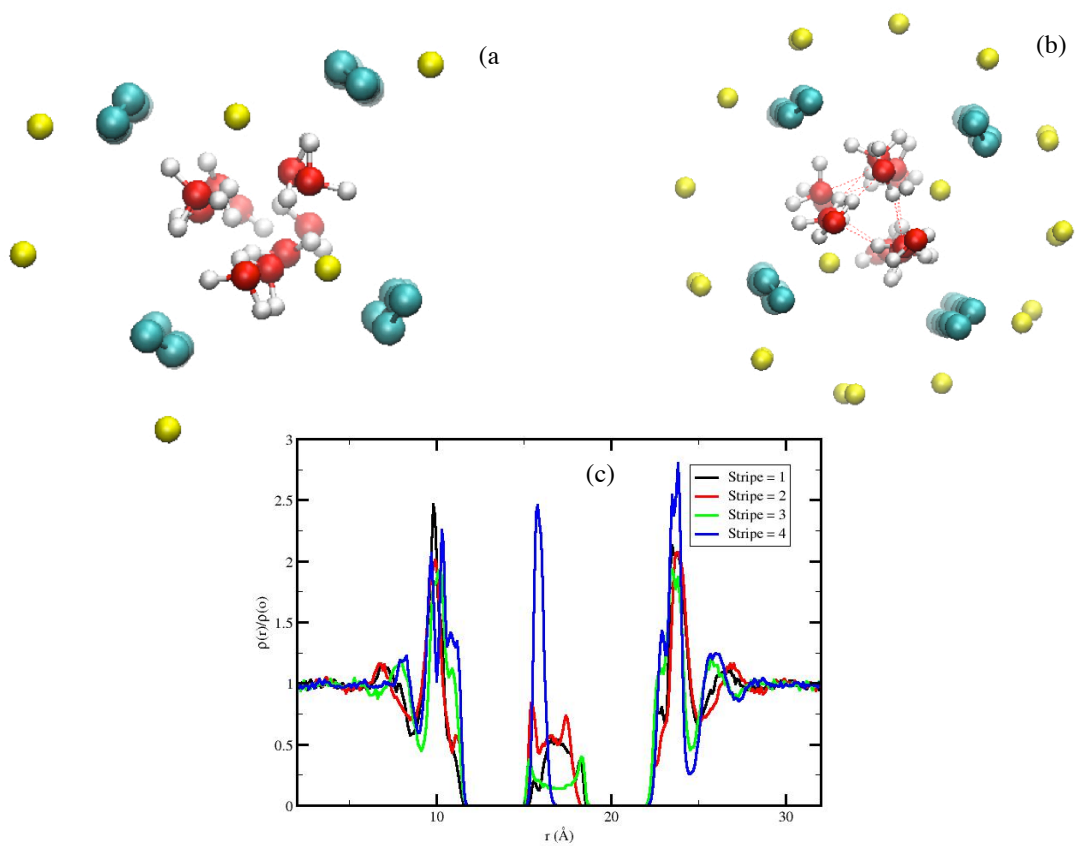

Fig. 6(a) Snapshot (top view) of simulation box having 48 charged carbon atoms at the terminal end of CNT, (b) Snapshot of simulation when same number of charged atoms are present at the central part of CNT (cyan show CNT walls, red oxygen atoms, white hydrogen atoms and yellow sodium ions), (c) radial density profile of water in the presence of stripes

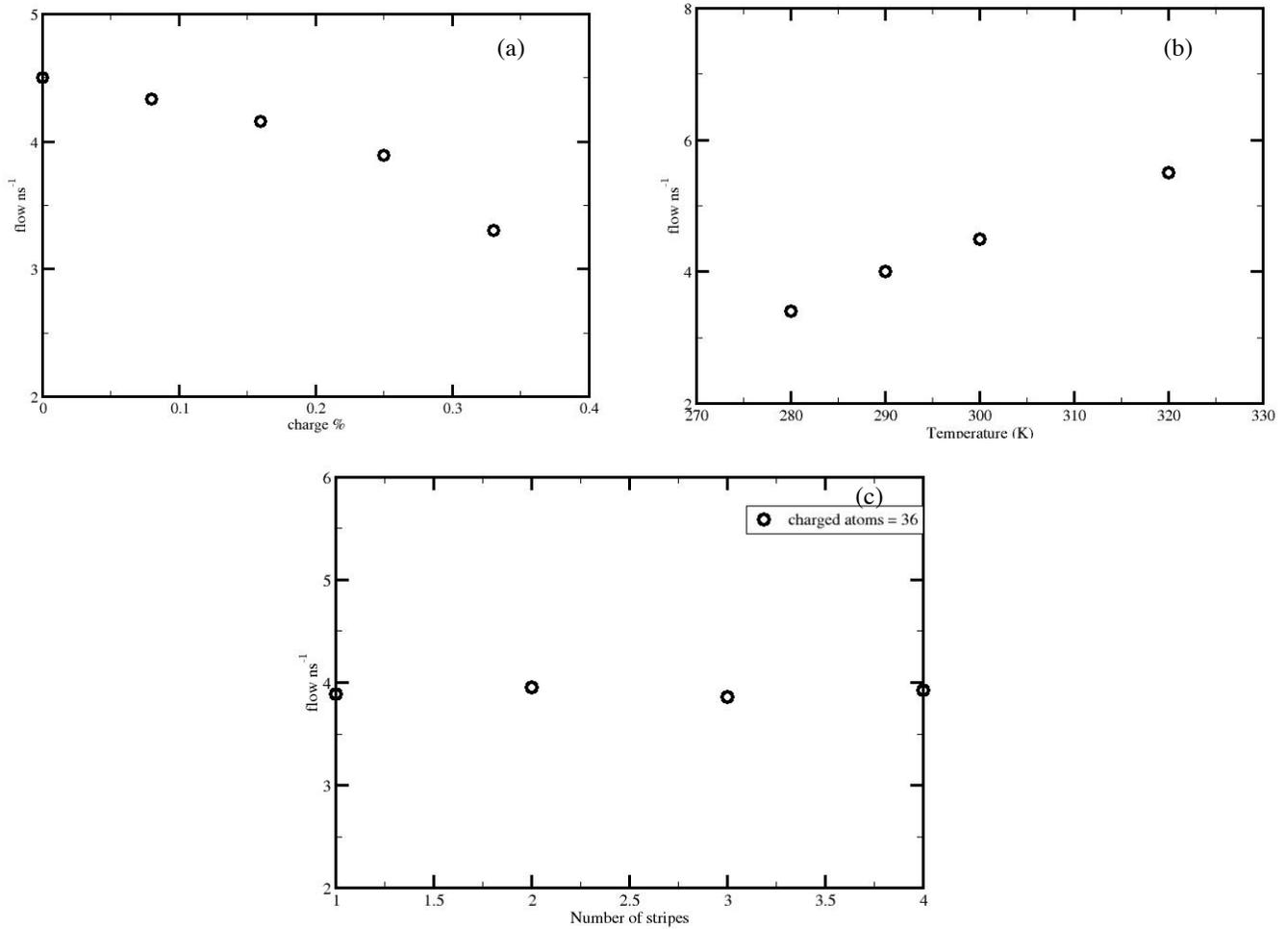

Fig. 7 (a) Effect of charge % on the flow of confined water, (b) Effect of temperature on the water flow (in pristine CNT), (c) Effect of stripes on the flow of water.

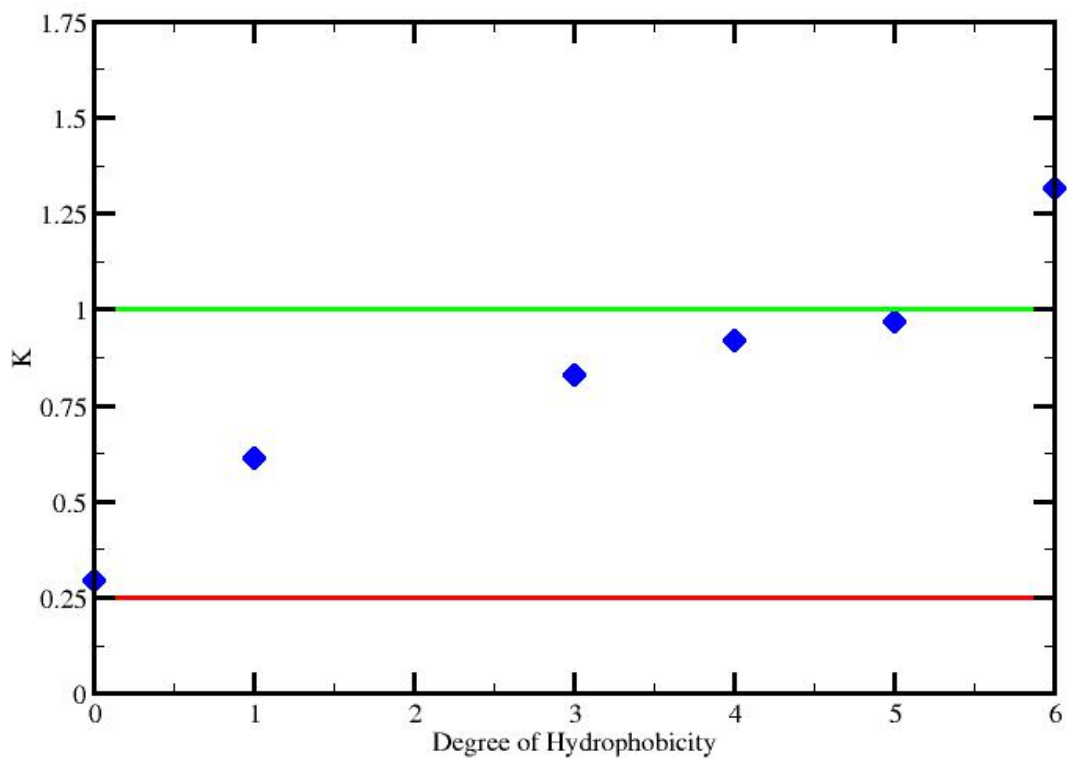

Fig. 8 Normalized variance as a function of hydrophobicity, horizontal lines indicates bulk water (0.25) and 1 for poissonian statistics. (Degree of hydrophobicity is inversely proportional to the charge density i.e. increase in hydrophobicity implies decrease in charge density. Here degree of hydrophobicity 6 implies pristine CNT).

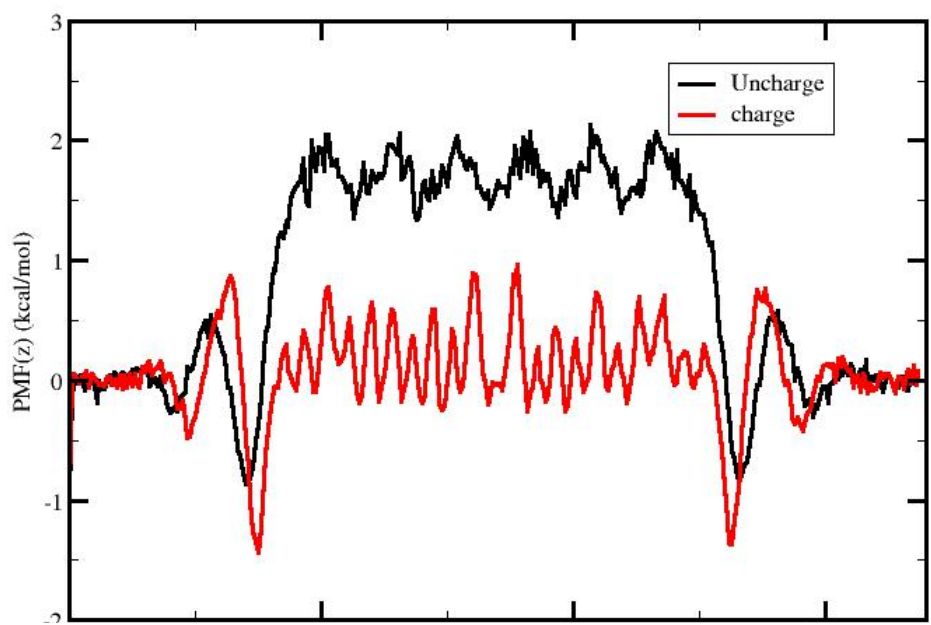

Fig. 9 Free energy profile of water permeating through pristine (block) and charged (red) CNT (charge = 48, at the central part of CNT)

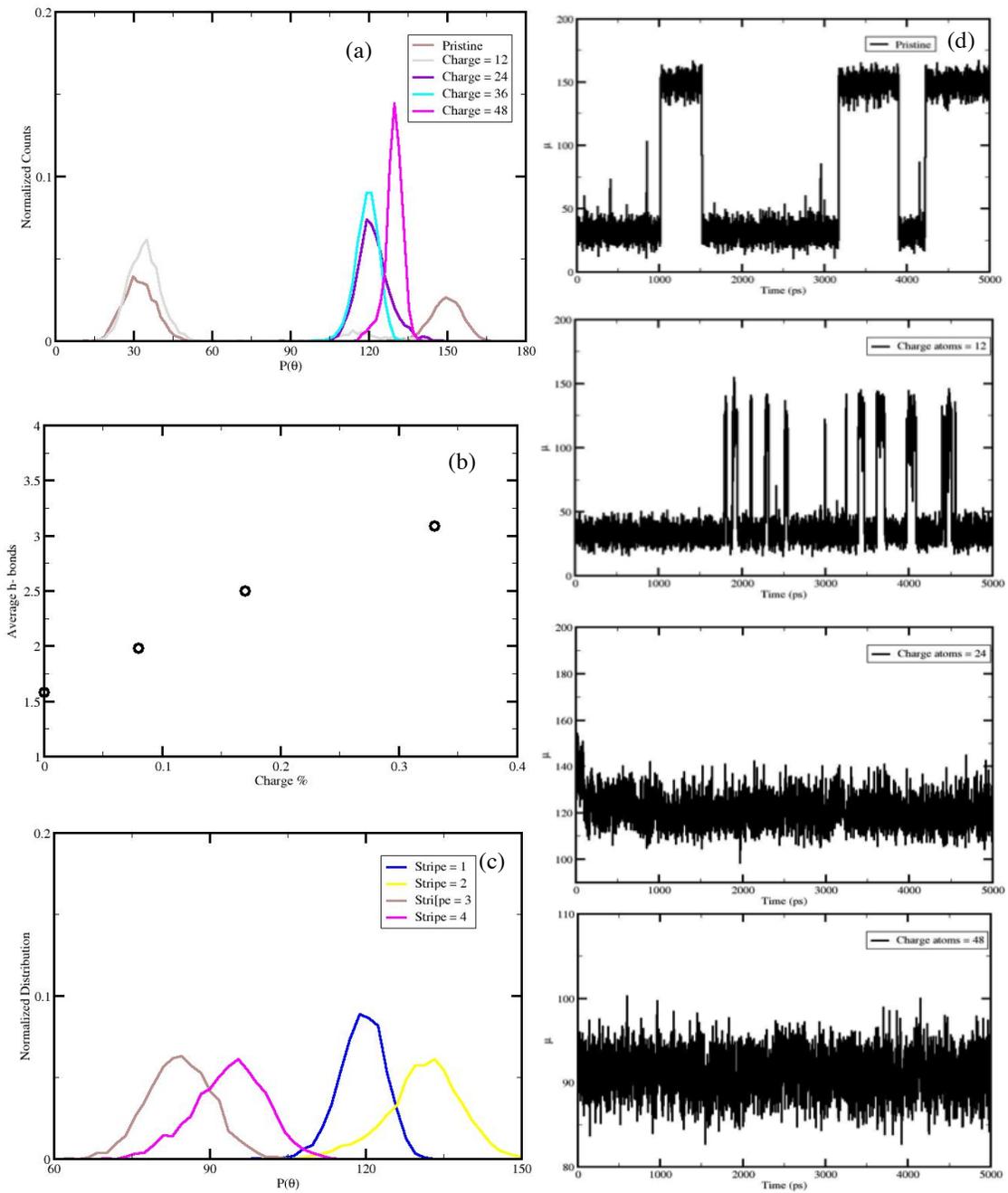

Fig. 10 (a) Change in the orientation angle by changing charge percentage, (b) Average number of hydrogen bonds in confined water, (c) Variation in orientation angle with stripes, (d) Flipping events in confined water inside different CNT.

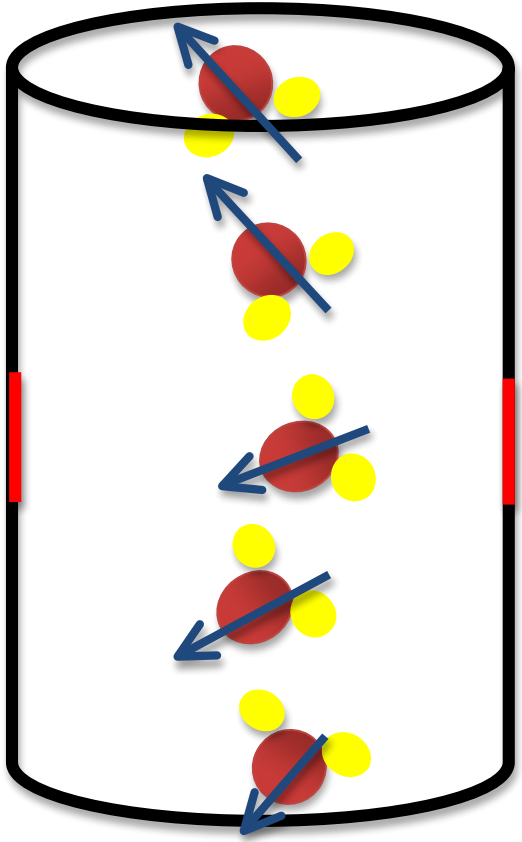

Fig. 11 Pictorial representation of changes in the dipole moment of confined water., black cylinder represents CNT, yellow & red balls represents hydrogen and oxygen respectively and the direction of dipole moment is shown by blue arrows.